# Enhanced superconductivity in the compressively strained bilayer nickelate thin films by pressure

Qing Li[1†], Jianping Sun[2,3†], Steffen Bötzel[4†], Mengjun Ou[1], Zhe-Ning Xiang[1], Frank Lechermann[4], Bosen Wang[2,3], Yi Wang[1], Ying-Jie Zhang[1], Jinguang Cheng[2,3*], Ilya M. Eremin[4*], Hai-Hu Wen[1*]

[1] National Laboratory of Solid State Microstructures and Department of Physics, Collaborative Innovation Center of Advanced Microstructures, Nanjing University, Nanjing 210093, China.

[2] Beijing National Laboratory for Condensed Matter Physics and Institute of Physics, Chinese Academy of Sciences, Beijing 100190, China.

[3] School of Physical Sciences, University of Chinese Academy of Sciences, Beijing 100190, China.

[4] Theoretical Physics III, Faculty of Physics and Astronomy Ruhr-University Bochum, D-44801 Bochum, Germany.

[†]These authors contribute equally to the work.
*Corresponding authors: jgcheng@iphy.ac.cn, Ilya.Eremin@ruhr-uni-bochum.de, hhwen@nju.edu.cn

**The discovery of high temperature superconductivity in the nickelate system has stimulated enormous interest in the community of condensed matter physics. Recently, superconductivity with an onset transition temperature ($T_c^{onset}$) over 40 K was achieved in $La_3Ni_2O_7$ and $(La,Pr)_3Ni_2O_7$ thin films at ambient pressure due to in-plane compressive strain. This**

observation has sparked enormous attention because measurements on superconducting properties can be accessible with many commonly used experimental tools. On the other hand, the $T_c$ in these thin films is much lower than that of the bulk bilayer nickelates under pressure. Here we report the enhancement of $T_c^{onset}$ to over 60 K by applying hydrostatic pressure on the compressively strained superconducting bilayer nickelate thin films. The $T_c^{onset}$ firstly ramps up with pressure, then it slightly drops down after reaching the maximum $T_c^{onset}$ at about 61.5 K under a pressure of 9 GPa, showing a dome-like phase diagram. Hall effect measurements reveal that the dominant charge carriers are hole-like with a slight enhancement of charge carrier density with pressure in accompanying with the increase of $T_c$. Our theoretical results demonstrate that the enhancement of $T_c$ arises from a cooperative amplification of magnetic fluctuations within and between the layers and increased metallicity under pressure. However, this enhancement exhibits saturation at higher pressures. These findings highlight the critical role of the interplay between interlayer and intralayer electronic correlations in bilayer nickelate superconductors and point to the potential of tuning $T_c$ through controlled manipulation of the electronic structure and interactions.

**Main text**

The discovery of superconductivity in bilayer nickelate $La_3Ni_2O_7$[1] with transition temperature ($T_c$) exceeding the boiling temperature of liquid nitrogen has drawn enormous attention in both theoretical[2-12] and experimental[13-26] studies in the past two years. The electronic configuration in bilayer nickelate system is $3d^{7.5}$ ($Ni^{2.5+}$) with dominantly the contributions of the $e_g$ orbitals, namely the $3d_{x^2-y^2}$ and $3d_{z^2}$ orbitals near the Fermi level[1,2-9]. This is strongly contrasted to the $3d^9$ configuration in cuprates[27] and infinite-layer nickelates[28]. Nowadays, nickelate superconductors are regarded as a new type of unconventional high-$T_c$ superconductor beside the cuprates and iron-chalcogenides[29,30]. However, the requirement of high-pressure conditions to achieve superconductivity in the bilayer system has inhibited the investigation of the superconducting mechanism and potential applications of nickelate superconductors. Therefore, the pursuit of superconductivity at ambient pressure in nickelates with the Ruddlesden-Popper (RP) phase has been an ongoing task, and several promising routes, such as chemical substitution or strain effect, have been proposed in previous studies[31-33].

Since the superconductivity in nickelate was first observed in an infinite layer thin film with $T_c$ about 9-15 K at ambient pressure[34], it is highly desired to explore ambient pressure superconductivity in $La_3Ni_2O_7$ system via strains or fine tuning of the oxygen stoichiometry. Indeed, thin films of bilayer nickelates on different substrates have already been successfully fabricated, showing a promising route to achieve ambient pressure superconductivity[35-37]. Very recently, signatures of superconductivity above 40 K in bilayer nickelate with a compressive strain (about -2 %) at ambient pressure is achieved in bilayer nickelate thin films grown on SrLaAlO$_4$ (SLAO) substrate[38,39] and the superconducting properties can be strongly enhanced through combined chemical substitution and precise optimization during growth and post oxidation treatment[39-41]. These findings have stimulated enormous interest, and several theoretical and experimental investigations have been carried out very quickly[42-46]. The angle-resolved photoemission spectroscopy (ARPES) measurements on this type of films reveal the presence of the $\alpha$ and $\beta$ Fermi surface sheets with the mixed $3d_{x^2-y^2}$ and $3d_{z^2}$ orbital content, but without consensus about the existence of $\gamma$ pocket which is purely derived from the $3d_{z^2}$ orbital[44,45]. The single particle tunneling spectroscopy measurements suggest the dominance of anisotropic s-wave gap(s) in the compressively strained bilayer nickelate films, supporting the Cooper-pairing scenario of bonding-antibonding $s^{\pm}$-wave superconductivity[46]. However, the maximum $T_c$ observed in bilayer nickelate thin films so far is much lower than that in bulk samples under high pressure. One possible scenario is that the large compressive strain in *ab* plane not only enhances the in-plane coupling to stabilize the ambient pressure superconductivity, but also increases the lattice constant in *c*-axis[38], thereby weakening the interlayer coupling along the *c* direction. In bulk La$_3$Ni$_2$O$_7$, the flat band ($\gamma$ band) with a large density of states just below the Fermi level with dominantly $3d_{z^2}$ orbital character may play the essential role for the occurrence of high-$T_c$ superconductivity[2,7,15,18]. Therefore, the $T_c$ in bilayer films might be further enhanced through chemical substitution or adding high pressure to shorten the *c*-axis lattice constant.

In this work, we report on the successful fabrication of superconducting bilayer heterostructure of $(La,Pr)_3Ni_2O_7/(La,Sr)_3Ni_2O_7$ on SLAO (001) substrate with *in-situ* ozone annealing. The $T_c^{onset}$ of the bilayer thin films is around 40 K, and the superconducting properties are pretty stable in air condition. By applying pressure, the $T_c^{onset}$ is monotonically enhanced up to 61.5 K at 9 GPa with zero resistance at about 5 K. With further increasing pressure, the $T_c^{onset}$ is saturated and decreases slowly with increasing pressure, giving a dome-shaped superconducting region. More importantly, the normal state transport measurements above $T_c^{onset}$ show a transition from metallic to weak insulating behavior with increasing pressure. Aided by the theoretical calculations, our results imply crucial effect of interplay of interlayer and intralayer multiorbital electronic correlations as well as the metallicity in governing the superconducting transition temperature.

According to the previous literatures, the superconducting bilayer nickelate thin films may have some diffusion of $Sr^{2+}$ in the substrate-film interface. Therefore, to enhance the stability of the superconductivity in the films, we firstly deposit an 1.5 unit-cell (UC) of $(La,Sr)_3Ni_2O_7$ layer on the $SrLaAlO_4$ (001) substrate, then deposit an 1.5 UC-$(La,Pr)_3Ni_2O_7$ thin film, forming an artificial structure as illustrated by the schematic plot in Fig. 1a. The quality of the obtained thin film was confirmed by X-ray diffraction (XRD) 2θ-ω scans along the (00*l*) direction. The data are shown in Fig. 1b and the XRD data of bare SLAO (001) substrate are also shown together. All the diffraction peaks from the films can be well indexed by the bilayer structure with the *c*-axis lattice constant of about 20.75 Å. This value is clearly elongated compared to the bulk samples[1,14], indicating that the film is compressively strained in plane by the substrate, but the *c*-axis lattice constant elongates a little bit. By fabricating the $(La,Pr)_3Ni_2O_7/(La,Sr)_3Ni_2O_7$ heterostructure ($LSP_{327}$) with a compressive strain and conducting precise *in-situ* ozone annealing, we can obtain superconducting nickelate thin films with bilayer structure. Interestingly, our films are quite stable in air, which may be induced by the specially designed heterostructure.

Figure 1d displays the temperature dependent resistivity of a variety of thin films. All the measured thin films exhibit good metallic behavior and clear superconducting transitions. The $T_c^{onset}$ varies from 28 to 40 K and most of the transitions have a resistivity drop over 90% at 2 K. We have also measured the $\rho(T)$ curves for $LSP_{327}$ films under different applied magnetic fields (Extended Data Fig. 1). The resistivity transitions are gradually pushed to lower temperatures by magnetic field as expected for a superconducting transition. The magnetic field dependent Hall resistance ($R_{xy}$) of $LSP_{327}$ thin film was measured between 40 and 200 K (Extended Data Fig. 2). As shown in Fig. 1c, the calculated Hall coefficient ($R_H$) is positive at measured temperatures and decreases with increasing temperature. Such behavior may indicate the multiband nature or correlation effect of the superconducting bilayer nickelate thin films in the normal state, and is consistent with the previous results on undoped $La_3Ni_2O_7$ and $La_2PrNi_2O_7$[38,40]. More importantly, we find that the superconducting transition of the grown films are very stable in air. For some films, after being placed in air for more than one month (see S1 and S2 in Fig. 1d), apart from a slight increase of the normal-state resistivity, the superconducting transition still looks very sharp without degeneration. The air-stable properties of the $LSP_{327}$ thin film samples enable us to conduct subsequent investigation on the transport properties of the bilayer nickelate thin films under pressure in different labs with different high-pressure cells.

As reported previously, the superconducting bilayer nickelate thin films were grown on a SLAO substrate with a compressive strain (about 2 % compared to bulk crystals), which consequently leads to abnormal elongation of the lattice parameter along the $c$-axis[1,14]. This will in turn lead to the elongation of out-of-plane Ni-O bonds, thereby weakening the interlayer $d_{z^2}$-orbital coupling. Therefore, one possible approach to increase $T_c$ in bilayer nickelate thin films is to reduce the $c$-axis lattice constant through pressure or chemical doping. Figure 2a presents the temperature dependence of resistance $R(T)$ curves of $LSP_{327}$ sample (S1) under various pressures up to 2.6 GPa measured in a piston−cylinder cell (PCC) with Daphne 7373 as the pressure transmitting medium (PTM). The normal state resistance gradually decreases with increasing pressure, and the residual resistance at 2 K decreases progressively. As shown in the inset of Fig. 2a, the zero resistance is reached at around 2.3 K above 2.0 GPa, indicating the better connection of the superconducting paths in LSP$_{327}$ thin films under pressure. The little step at about 173 K at ambient pressure may be due to an instant temperature instability in the warming up process. Figure 2b presents the enlarged view of the R(T) curves in the temperature region near the onset of superconducting transitions. Here, we define the onset of superconducting transition ($T_c^{onset}$) as the temperature where the resistance starts to drop and deviate from the linear extrapolation of the normal state data. It is clear that the $T_c^{onset}$ is monotonically enhanced from 37.1 to 45.3 K with increasing pressure up to 2.6 GPa, giving a slope dT$_c$/dP of about 3.15 K/GPa. This value is much larger than that reported in the infinite layer nickelate under high pressure[47]. We have also conducted the transport measurements on another thin film (S2) under high pressure and the results are presented in Extended Data Fig. 3. The similar evolution of R(T) curves and $T_c^{onset}$ under pressure indicates that the enhancement of superconductivity in the thin films is intrinsic and reproducible.

Figure 2c presents the full range R(T) curve of a LSP$_{327}$ thin film at 2.6 GPa. The line with red color is the fitting curve using $R(T) = R_0 + AT^n$ from T$_c$ to 100 K, and the fitting yields n = 1.85. In the temperature region above 200 K, the R(T) curve instead shows a T-linear dependence (n ~ 1) as indicated by the blue line. To see the evolution of the normal state transport behavior above T$_c$, we also performed the similar fitting on other $R(T)$ curves under different pressures and show the evolution of fitting parameter with pressure in Extended Data Fig. 4. The results indicate that the normal state transport behavior shows typical Fermi-liquid behavior ($n \sim 2$), but a slightly decrease of $n$ appears with increasing pressure, which indicates the gradual involvement of other exotic scattering or strange metal behavior under high pressures. Considering the strange metal behavior ($n$ is close to 1) is often observed in bulk bilayer nickelates under pressure[13,14,16] and infinite-layer superconducting films[48], we may expect that the normal state resistance would show a $T$-linear behavior when the $T_c$ approaches its optimal value. Pressure dependent Hall coefficient ($R_H$) measured at 40 and 60 K shows roughly a declining trend with increasing pressure as shown in Fig. 2d and Extended Data Fig. 5 and Extended Data Fig. 6, suggesting an increase of the charge carrier density. The slight increase of $R_H$ above 2.2 GPa may be induced by the solidification of PTM. Such behavior indicates that the carrier concentration increases upon applying a pressure. Higher carrier concentration is more conducive to the enhancement of superconductivity, which may also explain the enhancement of the superconducting transition temperature under pressure. Therefore, we may draw a conclusion that pressure will enhance the itinerancy of the charge carriers, leading to a better metallicity in the normal state.

To further study the evolution of $T_c$ in LSP$_{327}$ bilayer thin films under higher pressures, we perform high-pressure resistance measurements under various hydrostatic pressures up to 13 GPa using a cubic anvil cell (CAC) apparatus. This technique was successfully used in enhancing the superconducting transition $T_c$ of the infinite-layer nickelate thin films[47]. Figure 3a and Extended Data Fig. 7a show the $R(T)$ curves of two LSP$_{327}$ thin films (S3 and S4) measured with glycerol and Daphne 7373 as the liquid PTM, respectively. Although the temperature dependence of the normal-state resistance exhibits a drastic evolution, the superconducting onset transition temperature shows a clear enhancement at relative low pressures, then it saturates or slightly drops down at higher pressures. As can be seen in Fig. 3a, the room temperature resistance ($R_{300K}$) shows non-monotonic evolutions with increasing pressure, which exhibits a gradual decrease at $P$ < 7 GPa and then shows an anomalous increase above 7 GPa. At $P$ > 9 GPa, the normal state resistance above $T_c$ even shows an abnormal upward behavior in low temperature region. And the insulating feature becomes more and more obvious as the pressure further increases. For another sample S4 by using the Daphne 7373 as PTM (Extended Data Fig. 7), the $R(T)$ curves display nearly similar evolutions as sample S3 under high pressures, which first shows a decrease from 2 to 4 GPa and then increases gradually with further increasing pressure. The observed high-pressure results share highly similarity with the doped infinite-layer nickelate thin films that the defects or disorders dominate the electron scattering processes in resistance[47]. As is known, the delicate nickelate thin films can be easily and partially deteriorated at high pressures when the liquid PTM solidifies, which would produce disorders and defects in the thin film that will enhance the carrier scatterings. In the present study, we can see that the superconducting transition is sensitive to the pressure environment and still can remain visible up to 13 GPa by employing liquid PTM. This allows us to determine the superconducting onset transition temperature by using the method mentioned above.

In order to clearly track the evolution of the superconducting transition under high pressures, we vertically offset all $R(T)$ curves below 80 K for samples S3 and S4, as displayed in Fig. 3b and Extended Data Fig. 7b. We can see that the $T_c$ exhibits nonmonotonic evolution with increasing pressure. The definition of $T_c^{onset}$ is the same as that in Fig. 2b. For sample S3 and S4, the $R(T)$ shows $T_c^{onset} \approx 35.5$ K and $T_c^{onset} \approx 33$ K at ambient pressure, and the residual resistance at 1.5 K can be seen clearly, which should be ascribed to the small-sized sample (typical size of 0.6 mm × 0.3 mm) with prominent inhomogeneity for high-pressure measurements. Upon applying pressure to 2 GPa, we can observe perfect zero resistance with $T_c^{zero} \approx 2$ K and $T_c^{onset} \approx 45.1$ K for sample S3. With further increasing pressure, the superconducting transition temperatures of sample S3 increase to $T_c^{onset} \approx 52$ K and $T_c^{zero} \approx 3.5$ K at 3.5 GPa, and $T_c^{onset} \approx 58$ K and $T_c^{zero} \approx 5.3$ K at 5 GPa. With further increasing pressure, we can see that the zero-resistance state cannot be achieved above 7 GPa, but the thin film still exhibits relatively sharp transition even at 13 GPa. The optimal $T_c^{onset} \approx 61.5$ K at 9 GPa can be seen in the inset of Fig. 3b. In contrast, the superconducting transition is considerably broadened with a large residual resistance for sample S4, and it may be ascribed to the presence of substantial stress/strain induced by the solidification of Daphne 7373 upon compression and cooling down. However, even the transition becomes broad at high pressures above 9 GPa, we can still track the onset of superconducting transition, and its evolution is consistent with that of sample S3. In addition, the slight semiconducting like feature in the normal state above 9 GPa is intriguing. We are not sure whether it is induced by the solidification of the PTM, or it is intrinsic, since it occurs in the same way for the two runs of measurements with different PTM (and thus different solidification pressures/temperatures). This insulating feature was not observed in the previous studies on the pressurized effect on the infinite layer thin films[47], although the PMTs used are the same.

Figure 4 shows the pressure dependence of superconducting $T_c$ obtained from the high-pressure resistance measurements on LSP$_{327}$ thin films (S1-S4) with different high-pressure cells and PTMs. As can be seen, the phase diagram depicts explicitly a delicate evolution of superconducting transition, which shows clear difference to the results of La$_3$Ni$_2$O$_{7-\delta}$ bulk samples. With increasing pressure gradually, $T_c(P)$ shows pronounced positive pressure effect ($dT_c/dP \sim$ 3.5 K/GPa) and $T_c^{onset}$ exhibits a rapid increase from 35.5 K at ambient pressure to 61.5 K at 9 GPa and then decreases slowly to 55 K at 13 GPa, which features a dome-shaped superconducting phase with the maximum around 9 GPa. The optimal $T_c^{onset}$ and the critical pressure in the LSP$_{327}$ thin films are lower than the corresponding values in its bulk counterpart under a high pressure, which would imply a complex interplay of many factors concerning the interlayer and interlayer coupling, as well as the multiband effect in the system. This can get some reflections from our theoretical analysis as follows. In Fig. 4, we also mark the behavior of the normal-state resistance with different colors. Surprisingly, the maximum $T_c^{onset}$ corresponds with threshold pressure for the distinct normal-state behaviors very well, namely the metallic and weak insulating backgrounds. It is unlikely this weak insulating feature is induced by the solidification of the PTM since two different kinds of PTM have been used in different runs of measurements, but the data and the evolutions coincide quite well. In addition, the same measurement was taken to the infinite-layer thin film $Pr_{0.82}Sr_{0.18}NiO_2$ by using the same PTMs[47], where the normal state always shows a metallic behavior up to the highest pressure of about 12.1 GPa.

To get insight into the electronic origin of the $T_c$ evolution, we first perform density functional theory (DFT) calculations, where the application of pressure is modelled by the variation of the out-of-plane lattice constant $c$. To be specific, we compute the electronic structure for ambient pressure, $c_0$, and study additionally the cases of 1.5%, 3% and 5% reduced $c$ by using $c = c_0 - \Delta c$. In the next step we develop an effective low-energy model consisting of two $3d_{x^2-y^2}$ and $3d_{z^2}$ -orbitals located on two sites in a bilayer, which is obtained by maximally-localized Wannier down-folding[49]. The three-dimensional effective band structure is only shown for the ambient pressure, as well as the 5% case in Extended Data Fig. 8a. Three bands cross the Fermi level in our model independent of the applied pressure. They are denoted as α, β and γ bands, following the standard nomenclature[2]. The α and β bands are of mixed $3d_{x^2-y^2}$ and $3d_{z^2}$ orbital character with dominant contribution of the former near the Fermi level. In contrast, the more flattish γ band is of $3d_{z^2}$ orbital character near the Fermi level. To include the effects of the electronic correlations, we perform the rotationally invariant slave boson (RISB) approximation, using $U$=7 eV, $V$=1 eV[50]. The renormalized band structures for each pressure are shown in Extended Data Fig. 8b and the resulting Fermi surface for ambient and 3% reduced lattice constant *c* are shown exemplarily in Fig. 5d. Observe that the largest effect of the pressure for the Fermi surface and the bandwidth occurs for the γ pocket. This is expected due to the strongest effect of the *c*-lattice variation on the $3d_{z^2}$ orbital. The γ band shifts upwards with respect to the Fermi level and increases its bandwidth (see also Extended Data Fig. 8). This upwards shift corresponds to a charge flow from $3d_{z^2}$ to $3d_{x^2-y^2}$ orbital with increasing pressure and reflects the reduction of the crystal field splitting (for $\Delta c/c_0$ = 0, 1.5, 3, 5%: $\Delta_{\text{cf}}$ = 541, 391, 316, 184 meV). The crystal field splitting is also influencing the leading instability in the linearized gap equation based on RPA spin calculations[51].

In addition, there is an overall increase of itinerancy of the bands, which is again due to stronger hybridization of $3d_{x^2-y^2}$ and $3d_{z^2}$- orbitals and a charge flow from $3d_{z^2}$ to $3d_{x^2-y^2}$ orbital upon pressure. The increased metallicity is illustrated in Fig. 5d where we plot the quasiparticle weights (residue) of the two orbitals for various pressures. This is consistent with the Hall coefficient data, shown in Fig. 2d indicating stronger metallicity of the normal state upon applied pressure.

Next, we perform calculations of the spin susceptibility as described in the Extended Data Fig. 9 and Extended Data Fig. 10, and the calculated bare spin susceptibility is depicted in Figs 5a and 5b. Both, even (in-plane) and odd (out-of-plane) spin components are shown $\chi^e = (\chi_{bb} + \chi_{aa})/2$ and $\chi^o = (\chi_{ab} + \chi_{ba})/2$[52,53]. Here, for example, the in-plane $\chi_{bb}$ describes contributions due to the scattering from a bonding to a bonding band. The odd channel essentially describes the scattering between bonding and antibonding bands. Consequently, strongly dominant spin fluctuations in the odd channel do naturally provide the pairing glue for the bonding-antibonding $s_\pm$-wave. An enhancement with decreasing *c* in the odd channel is seen, when the slave boson quasiparticle weights, which are shown in Fig. 5c, are included. Note that the renormalization for the $3d_{z^2}$ orbital is stronger than that for the $3d_{x^2-y^2}$ orbital. The difference in the orbital renormalization, however, is reduced through reduction of *c*-axis lattice constant. However, the even spin fluctuations are also significantly sensitive to the changes of *c* lattice constant. Especially, a dome like behavior is seen for the $q_1^e$ and $q_3^e$ peaks, which is consistent with the experimental observation shown in Fig. 4.

Overall, we see that the effect of the pressure modifies both the in-plane and the out-of-plane magnetic response, which makes it challenging to single out the main effect responsible for the increase of $T_c$ with pressure and even make further assessment on the most probable symmetry of the superconducting gap. We remind that the main candidate solutions for the superconducting gap in the bilayer nickelates are either bonding-antibonding $s_\pm$-wave or d-wave (either $d_{xy}$ or $d_{x^2-y^2}$) gap structures[4,51]. At the same time, we can identify universal features in the electronic response upon the application of pressure, which are (i) increased hybridization between $3d_{z^2}$ and $3d_{x^2-y^2}$ orbitals (ii) larger itinerancy of the pressurized system, and (iii) enhanced antiferromagnetic spin fluctuations both in the in-plane and out-of-plane channels. Moreover, these effects quickly saturate upon larger pressures explaining why no further increase of $T_c$ and even its reduction is observed. To see that these universal features stay the same for the slight variation of the electronic structure we repeat the same calculations by doping of 0.1 holes or 0.2 holes per Ni atom assuming potential doping effect of the substrate, as shown in Extended Data Fig. 9 and Extended Data Fig. 10. Remarkably, the universal features we observe for non-doped compound stay the same.

The combination of strain engineering and high pressure is an essential tool in the research of nickelate superconducting thin films. For example, in infinite-layer thin films, the $T_c$ can be monotonically enhanced above 30 K at 12.1 GPa[47]. Now in the compressively strained bilayer nickelate thin film, the $T_c^{onset}$ exhibits a dome-like phase diagram. Our theoretical calculations show that this is likely a synergic effect of increased hybridization between $3d_{z^2}$ and $3d_{x^2-y^2}$ orbitals, resulting in the charge flow between the orbitals and larger itinerancy of the pressurized system, together with the enhanced antiferromagnetic spin fluctuations both in the in-plane and out-of-plane channels. Moreover, these effects quickly saturate upon larger pressures explaining why no further increase of $T_c$ and even its reduction is observed. Our combined experimental and theoretical efforts highlight the critical role of the interplay between interlayer and intralayer electronic correlations in bilayer nickelate superconductors, together with the delicate manipulation of the multiband contributions. This will point to the potential of tuning $T_c$ through controlled manipulation of the electronic structure and interactions.

In summary, we report the synthesis of air-stable superconducting bilayer nickelate thin films and investigate the effect of pressure on the superconducting properties of the bilayer nickelate thin films. At ambient pressure, the $T_c^{onset}$ of the as grown thin films is approximately 35-40 K, and the superconducting transitions are pretty stable in air. Under high pressures, the $T_c^{onset}$ is monotonically enhanced up to 61.5 K at 9 GPa and decreases slowly with further increasing pressure, giving a dome-like shape of the superconducting phase. Theoretical calculations clearly indicate the significant role played by the cooperative amplification of magnetic fluctuations within and between the layers and increased metallicity of the pressurized compounds in enhancing $T_c$.

*Note added:*

During the preparation of our manuscript, we realize an independent high pressure study on La$_3$Ni$_2$O$_7$ thin films grown on different substrates without superconducting transition at ambient pressure[54]. Under high pressures, the non-superconducting La$_3$Ni$_2$O$_7$ films shows superconducting transition and the reported maximum $T_c^{onset}$ is similar to our present work. Our work here provides a more detailed demonstration of superconductivity of the compressively strained thin films with already superconductivity, and a comprehensive theoretical understanding.

## Method

The nickelate thin films with bilayer structure were grown by pulsed laser deposition (PLD) technique using (001)-oriented $SrLaAlO_4$ single crystal as substrate. The growth process and parameters were similar to that reported previously[46]. In order to fabricate bilayer heterostructure of $La_{2.5}Pr_{0.5}Ni_2O_7$/$La_{2.85}Sr_{0.15}Ni_2O_7$, we synthesized the nominal stoichiometric targets by sol-gel method. The growth process involved the deposition of the $La_{2.85}Sr_{0.15}Ni_2O_7$ film first, followed by the deposition of $La_{2.5}Pr_{0.5}Ni_2O_7$ film in an environment with oxygen partial pressure of 30-35 Pa. The as grown thin films have already shown some weak superconducting transition at low temperatures. To achieve better superconductivity, the post-treatment was performed in the growth chamber at 550 °C with an annealing time of 0.5–2 hours and an ozone pressure (~7 wt% under 20-25 Pa) and flow rate of 10-25 sccm. The X-ray diffraction (XRD) 2θ-ω scans was obtained using a Bruker D8 Advanced X-ray diffractometer with a Cu $K_{α1}$ radiation source ($λ$ = 1.541 Å). The crystal structure was modeled using VESTA software[55]. The temperature-dependent electrical resistance measurements at ambient pressure were carried out on a physical property measurement system (PPMS, Quantum Design) using the standard four-probe methods.

Standard four-probe method was employed to measure temperature-dependent resistance of the $LSP_{327}$ thin films with the electric current applied within the ab-plane. High pressure (<2.6 GPa) resistance measurements were carried out in a PCC apparatus (HPC-33, Quantum Design). The sample was immersed in a liquid pressure transmitting medium of Daphne 7373 in a Teflon cell. Hydrostatic pressures were generated by a nonmagnetic BeCu/NiCrAl piston-cylinder cell. The applied pressure at low temperature was determined by the shift in $T_c$ of a high quality Pb single crystal. The Hall resistance under high pressures was measured using a five-probe method. For measurements at higher pressures, we employ a palm-type CAC to measure its resistance under various pressures up to 13 GPa with the glycerol as the liquid PTM, and to 12 GPa with the Daphne 7373 as the liquid PTM. The pressure values inside the CAC were estimated from the pressure-loading force calibration curve pre-determined by measuring the Bi and Pb at low temperatures.

To get insights into the electronic states of the thin films, DFT in the local density approximation within a mixed-basis pseudopotential framework is applied. The application of pressure is modelled by variation of the out-of-plane lattice constant and fixed in-plane lattice constant within a bulk-like setting. To be specific, we use $c_0$ for ambient pressure and study additionally the cases of 1.5%, 3% and 5% reduced $c$ by using $c = c_0 - \Delta c$.

In bilayer systems, the electronic states in band space can be approximately diagonalized by introducing bonding and antibonding states $c_{b/a,k} = \frac{1}{\sqrt{2}}(c_{1,k} \pm c_{2,k})$, where $c_{1,k}$ and $c_{2,k}$ denote the annihilation operators for an electron in layer 1 and 2 with momentum $k$, respectively. Within the description of the two orbital bilayer model, the $\alpha$ and $\gamma$ bands belong to the bonding subspace whereas the β band is of antibonding character[2]. Scattering events can be classified in those which happen within the bonding or antibonding bands and those which happen between both subspaces. The even and odd susceptibilities are defined by $\chi^{e/o} = 2\chi_{11,11} \pm 2\chi_{11,22}$, where orbital and spin indices are omitted. The physical spin susceptibility in the paramagnetic state is determined by even and odd channels as

$$\chi^{\text{spin}}(q) = \sum_{\mu_1\mu_2}[\chi^e_{\mu_1\mu_1,\mu_2\mu_2}(q)\cos^2\left(\frac{q_z d}{2}\right) + \chi^o_{\mu_1\mu_1,\mu_2\mu_2}(q)\sin^2\left(\frac{q_z d}{2}\right)],$$

where $q = (\boldsymbol{q}, \omega)$ and the sum $\mu_{1,2}$ run over both $\text{Ni} - e_g$ orbitals[52,53]. The bare susceptibility is calculated using the renormalized band structure. Furthermore, the slave-boson quasi-particle weight is dominated by the diagonal part. These diagonal elements $Z_\mu$ can be used as prefactors to account for the orbital selective quasi particle weights: $\sqrt{Z_{\mu_1}Z_{\mu_2}Z_{\mu_3}Z_{\mu_4}}\chi^{e/o}_{\mu_1\mu_2,\mu_3\mu_4}(q)$[56]. While the spin susceptibility is only shown along the high-symmetry path in the $q_z = 0$ plane, the full $k_z$ dependence of the band structure is included in the calculation. To be specific, a $200 \times 200 \times 25$ $\boldsymbol{k}$-mesh and bulk filling has been used to obtain the results in Fig. 5. To account for the possibility of hole doping, the calculations were repeated for 0.1 and 0.2 holes per Ni atom. The results are shown in Extended Data Fig. 9 and Extended Data Fig. 10, respectively. The discussed trends are similar for all doping.

## Data availability

All data needed to evaluate the conclusions in the paper are present in the paper. Additional data related to this paper may be requested from the authors.

## Acknowledgments

We thank the useful discussions with Daoxin Yao, Meng Wang and Matthias Hepting. We appreciate the kind help in the polishing of the thin films given by Zhe Liu. This work was supported by the National Key R&D Program of China (Grant No. 2022YFA1403201 and No. 2023YFA1406100), National Natural Science Foundation of China (Grant No. 12494591, No. 11927809, No. 12025408, No. 12174424, No. 12494592, No.123B2055 and No. 12204231), the CAS Project for Young Scientists in Basic Research (2022YSBR-048), the Youth Innovation Promotion Association of CAS (2023007).

## Author contributions

The nickelate thin films were grown by M.O., Y.W., Q.L., and H.-H.W. The high-pressure electrical transport measurements in PCC were performed by Z.-N.X., Y.-J.Z. and Q. L. with assistance from H.-H.W. High-pressure resistance measurements in CAC were done by J.S., B.W and J.C. Theoretical calculations and their analysis were carried out by S.B., F.L and I. E. Q. L., I.E. and H.-H.W. analyzed the experimental data and wrote the manuscript with the inputs from all co-authors.

## Competing interests

The authors declare that they have no competing interests.

# Figures

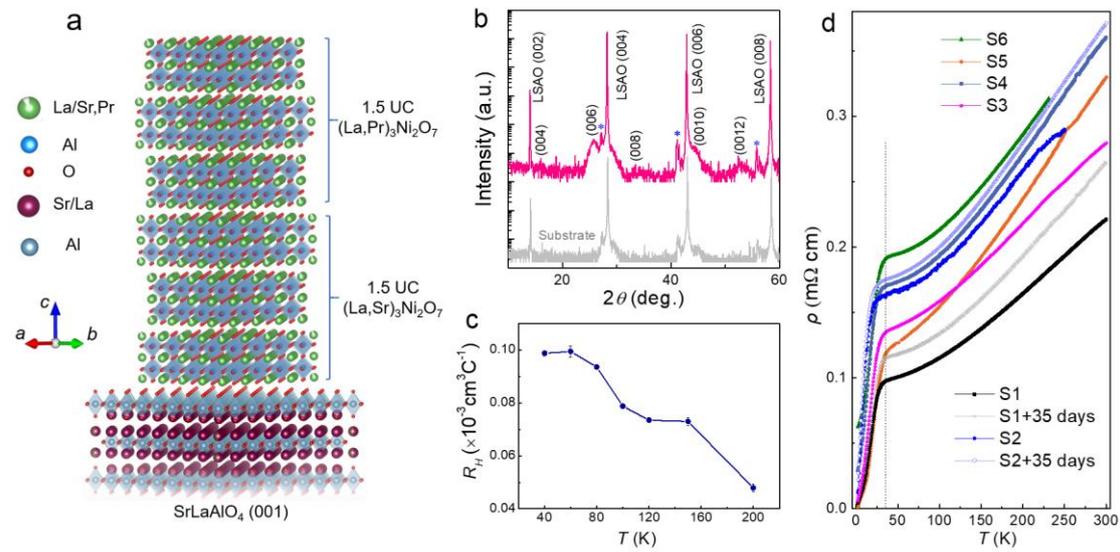

**Fig. 1 | Characterization of the superconducting thin films at ambient pressure. a,** Schematic crystal structure of the $(La,Pr)_3Ni_2O_7/(La,Sr)_3Ni_2O_7$ heterostructure grown on SLAO (001) substrate. The 3-UC bilayer structure is accordingly shown for the heterostructure of 1.5-UC $(La,Sr)_3Ni_2O_7$ and 1.5-UC $(La,Sr)_3Ni_2O_7$. **b,** X-ray diffraction 2θ-ω symmetric scans of the 7-nm thick $LSP_{327}$ thin film (upper) grown on SLAO (001) substrate (bottom). **c,** Normal state Hall coefficient ($R_H$) as a function of temperature measured at ambient pressure. The data were collected on sample S6. **d,** The $\rho(T)$ curves for different $LSP_{327}$ films and the time dependence at ambient pressure. The superconducting transition is clearly seen with the $T_c^{onset}$ ranging from 28 to 40 K, and it remains basically unchanged over one month for S1 and S2.

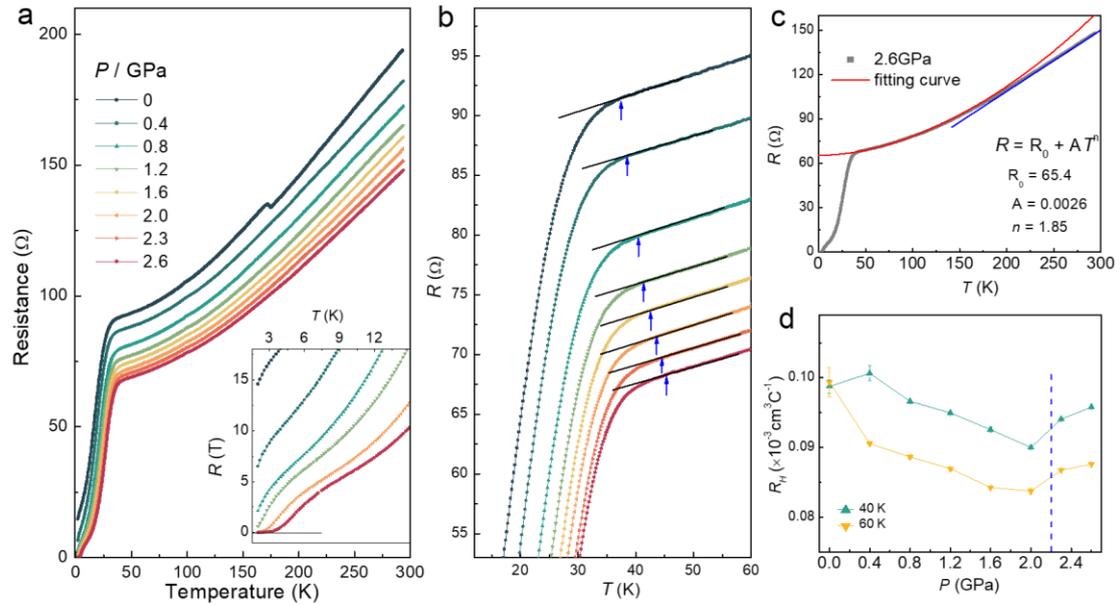

**Fig. 2 | Pressure dependent $R(T)$ curves and transport properties in PCC. a,** Temperature dependence of resistance of LSP$_{327}$ thin film (S1) under various pressures up to 2.6 GPa with Daphne 7373 as the PTM. **b,** Enlarged view of $R(T)$ curves in the temperature region from 12 to 60 K. The black lines and blue arrows illustrate the definition of the onset transition of superconductivity. **c,** The full temperature range $R(T)$ curve of LSP$_{327}$ thin film at 2.6 GPa. The temperature dependent resistance above $T_c^{onset}$ can be well fitted by the equation $R(T) = R_0 + AT^n$ with $n$ = 1.85 (red line). The blue line indicates linear fit of the $R(T)$ curve above 200 K. **d,** The evolution of Hall coefficients ($R_H$) under different pressures at 40 and 60 K measured on another sample (S6). The vertical blue dashed line indicates the solidification pressure of Daphne 7373 at room temperature.

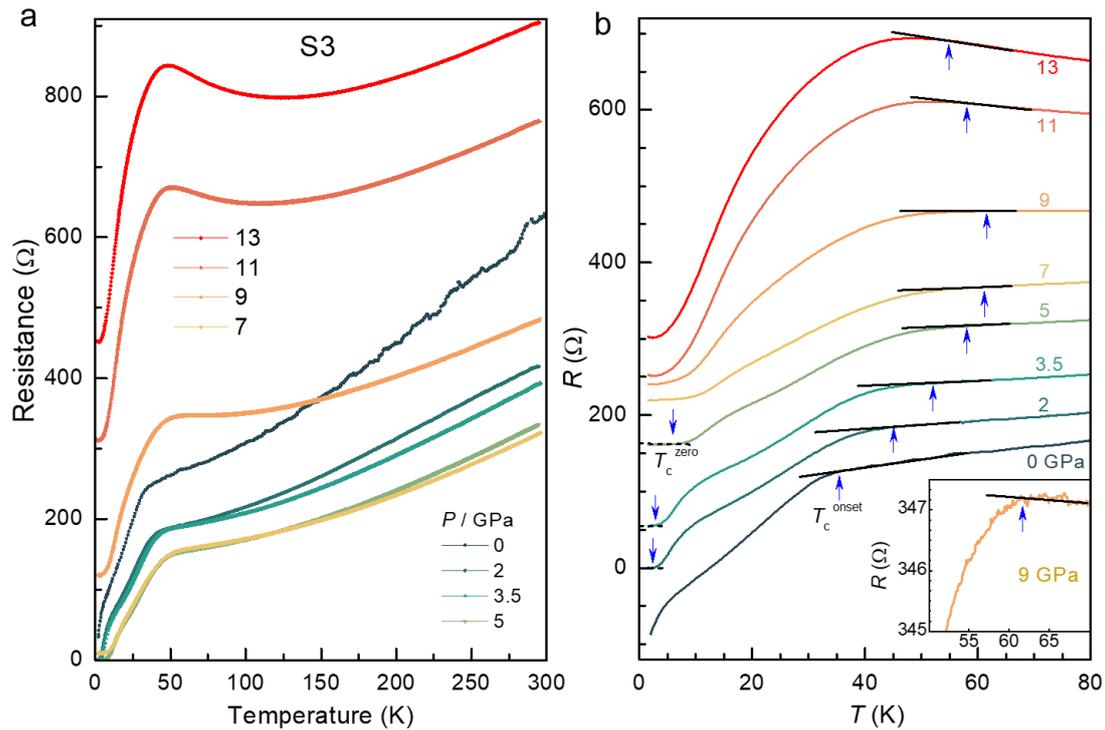

**Fig. 3 | Pressure dependent $R(T)$ curves in CAC. a,** Temperature dependence of resistance $R(T)$ of a LSP$_{327}$ thin film (S3) under various pressures up to 13 GPa with glycerol as PTM. **b,** The enlarged view of $R(T)$ curves below 80 K for S3. To illustrate the variation of the superconducting transition temperatures with pressure more clearly, the $R(T)$ curves in **b** have been vertically shifted. The $T_c^{onset}$ (up-pointing blue arrow) was determined as the temperature where resistance starts to deviate from the extrapolated normal-state behavior and the $T_c^{zero}$ (down-pointing blue arrow) was defined as the temperature when the resistance reach zero. The optimal $T_c^{onset} \approx 61.5$ K at 9 GPa can be seen in the inset of **b**.

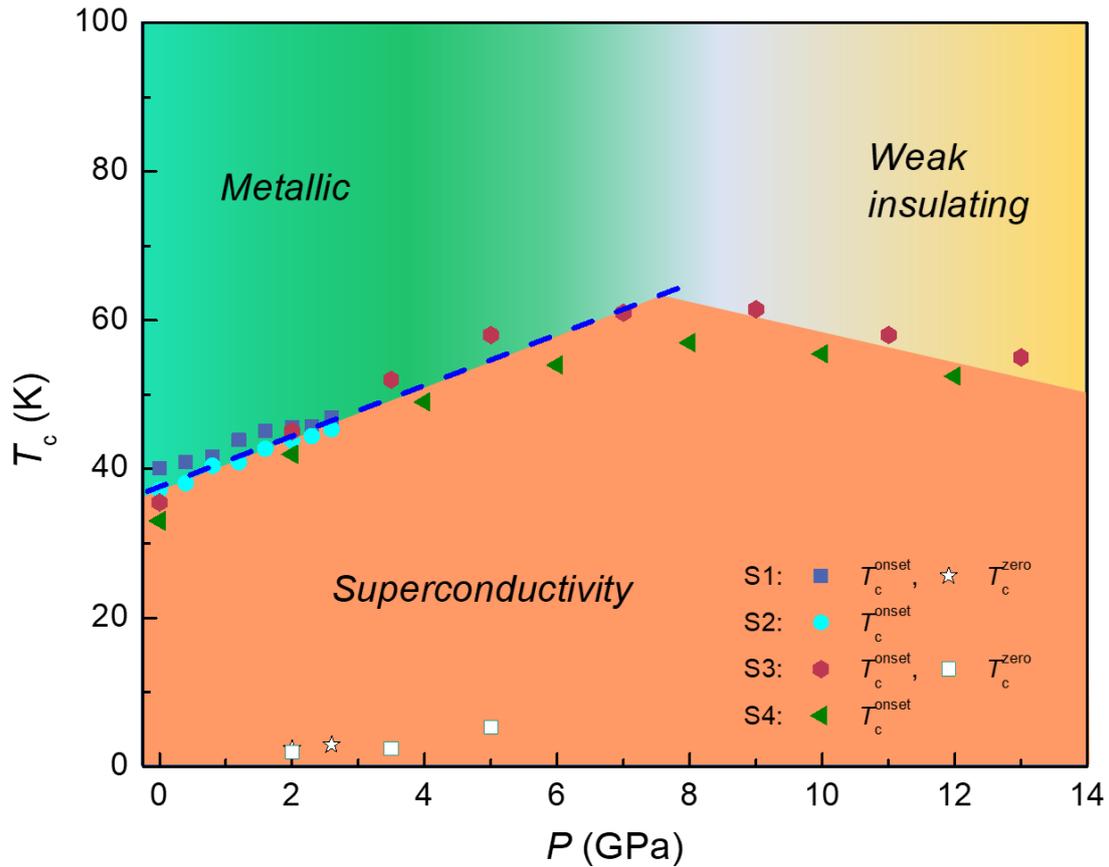

**Fig. 4 | The phase diagram of superconducting transition temperature with pressure.** The pressure dependence of the superconducting transition temperatures $T_c^{onset}$ and $T_c^{zero}$ were determined from the $R(T)$ measurements. The blue dashed line indicates the linear fit to the $T_c^{onset}$ below 7GPa, giving the ratio of $dT_c/dP$ of about 3.5 K/GPa. After about 9 GPa, the transition temperature shows a saturation and then a slow declining. The green and yellow region represents the metallic and weak insulating behavior above $T_c$ under high pressures.

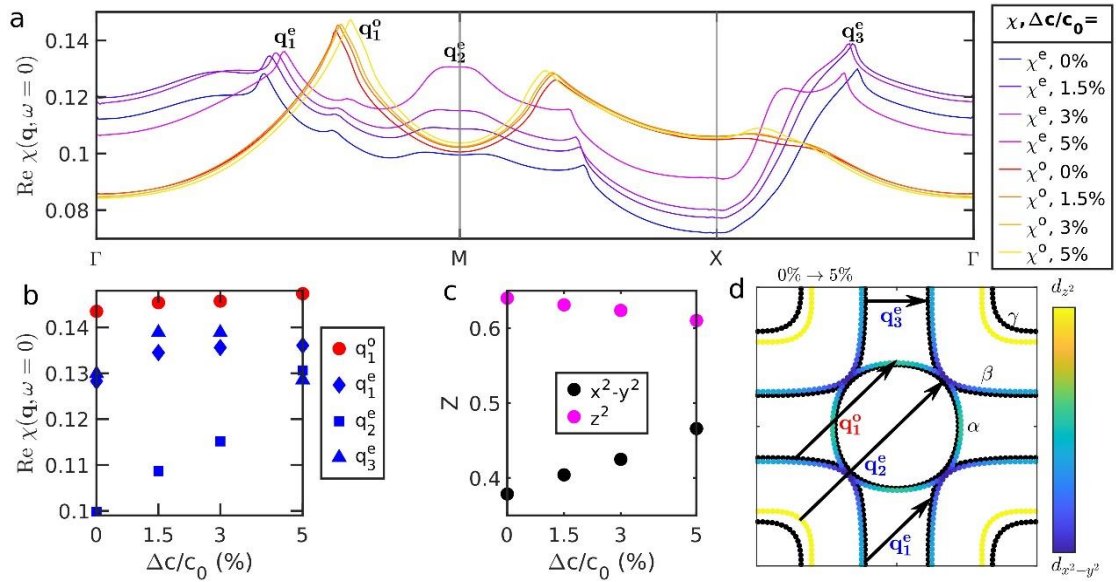

**Fig. 5 | Theoretical evolution of the spin fluctuations upon pressure. a,** In-plane (even) and out-of-plane (odd) channels of the bare static spin susceptibility using the slave boson renormalized band structure and slave boson quasiparticle weights for different values of the out-of-plane lattice constant $c$. **b,** Evolution of the main peak values of the susceptibility shown in **a** with the change of the lattice constant $c$. **c,** Calculated evolution of the slave boson quasiparticle weights for different orbitals. **d,** Calculated Fermi surface evolution for ambient pressure and smallest applied pressure. Arrows indicate the scatterings associated with the peaks in **a** and **b**.

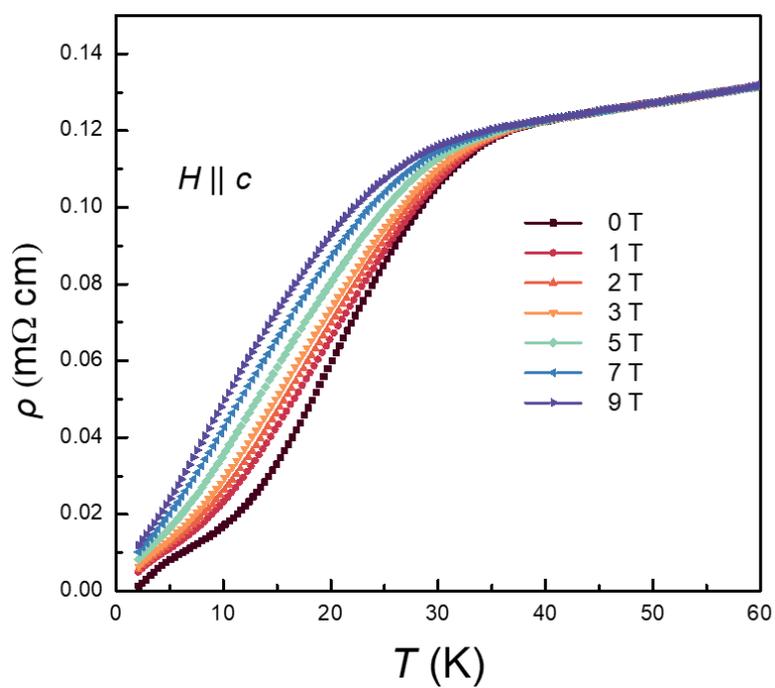

**Extended Data Fig. 1 | Temperature dependent resistivity for an LSP$_{327}$ thin film (S5) under various magnetic fields applied along the *c*-axis.**

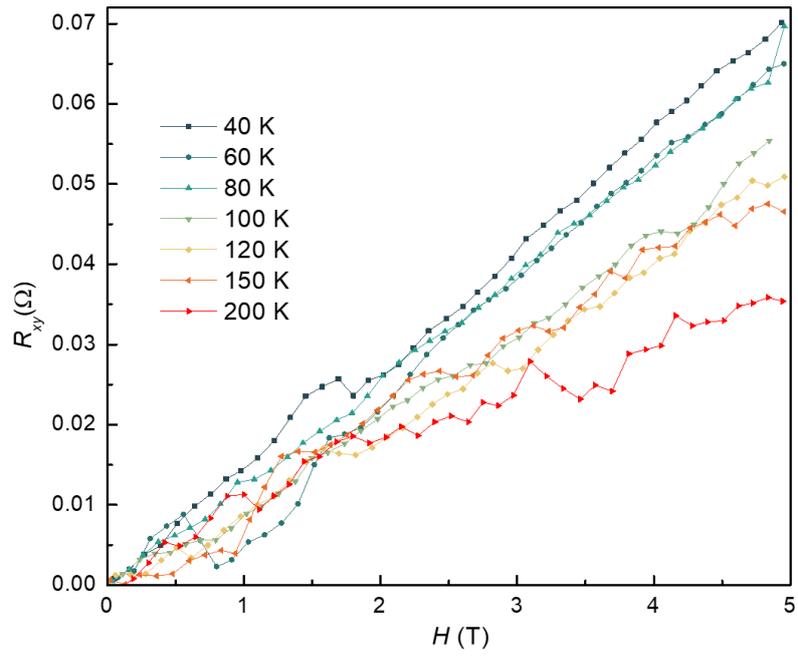

**Extended Data Fig. 2 | Magnetic field dependent Hall resistance ($R_{xy}$) at different temperatures for an LSP$_{327}$ thin film (S6) at ambient pressure.**

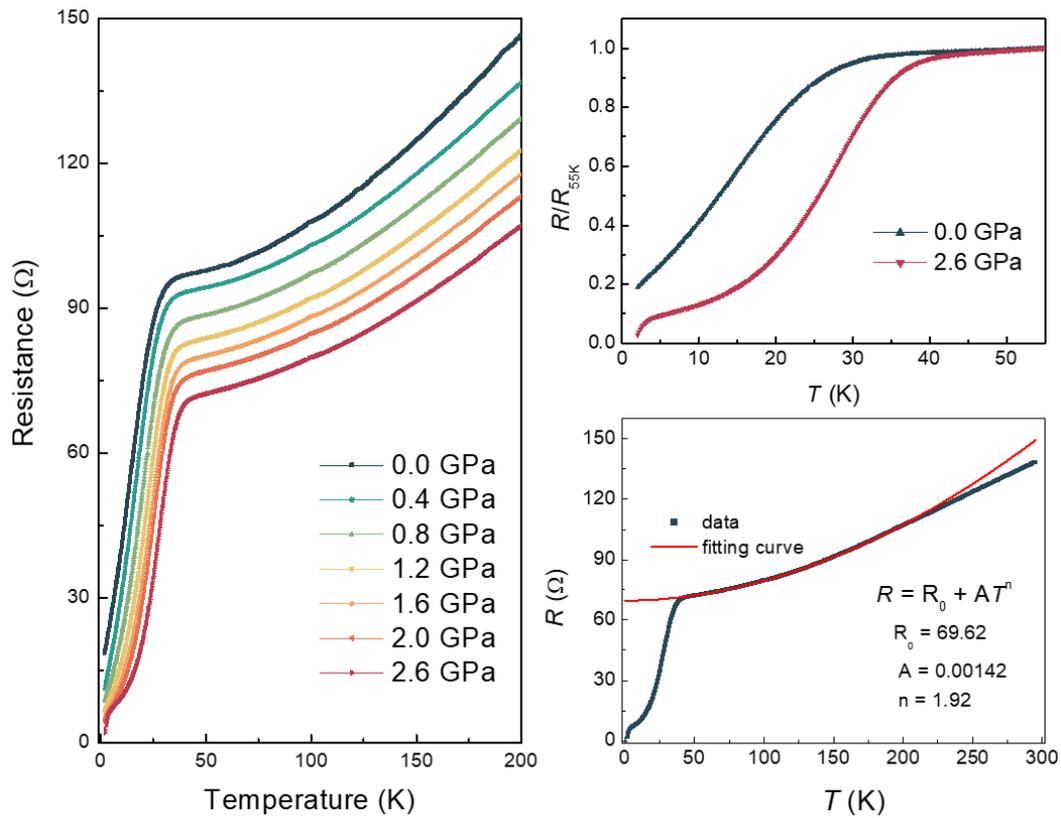

**Extended Data Fig. 3 | Pressure dependent $R(T)$ curves for S2 in PCC. a,** Temperature dependence of resistance of another LSP$_{327}$ thin film (S2) under various pressures a up to 2.6 GPa with Daphne 7373 as the PTM. **b,** Normalized $R/R_{55K}(T)$ curves in the temperature region from 2 to 55 K at 0 and 2.6 GPa. The superconducting transition temperature is clearly enhanced under pressures. **c,** The $R(T)$ curve of LSP$_{327}$ at 2.6 GPa. The temperature dependent resistance from $T_c$ to 100 K can be well fitted by the equation $R(T) = R_0 + AT^n$ with $n$ = 1.92. With temperature above 200 K, the fitting curve deviates clearly from the experimental data.

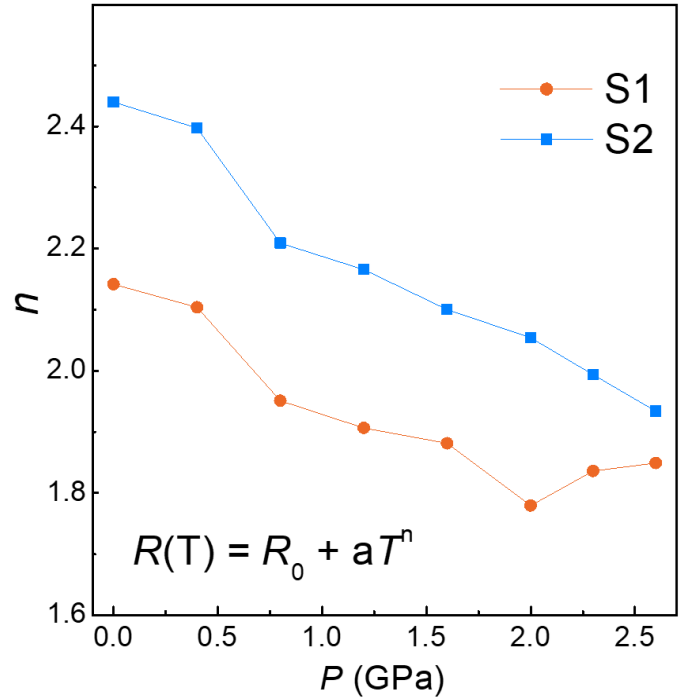

**Extended Data Fig. 4 | Pressure dependence of the exponent *n* that derived from the fitting to the *R(T)* data of S1 and S2 with the formula $R(T) = R_0 + AT^n$ in the temperature region from $T_c$ to 100 K.** As the pressure increases, the *n* value gradually decreases.

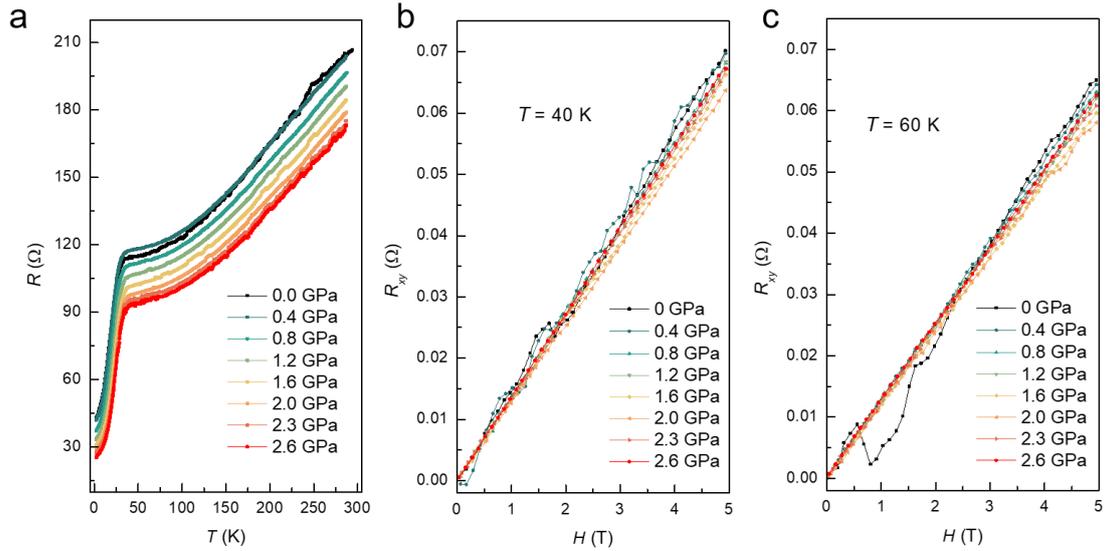

**Extended Data Fig. 5 | Temperature dependence of resistance and Hall effect measurement for S6. a,** Temperature dependence of resistance of another LSP$_{327}$ thin film (S6) under various pressures up to 2.6 GPa. **b, c** Magnetic field dependent Hall resistance ($R_{xy}$) under pressures at 40 and 60 K, respectively. the slop of Hall resistance versus magnetic field curves at 2.3 and 2.6 K show some abnormal increases, which may be caused by the solidification of PTM Daphne 7373 above 2.2 GPa.

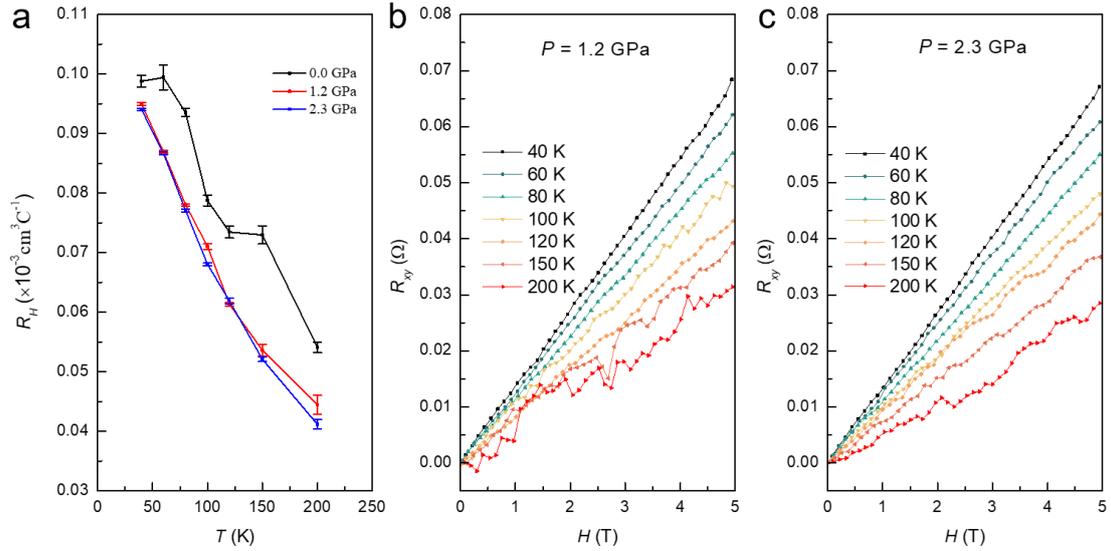

**Extended Data Fig. 6 | Raw data of Hall effect and coefficient. a,** The evolution of Hall coefficients ($R_H$) under different temperature under various pressures. **b, c** Magnetic field dependent Hall resistance ($R_{xy}$) under different temperatures at 1.2 and 2.3 GPa, respectively. Under high pressures, the $R_H$ still shows clear temperature dependence and the charge carrier is hole type in the whole temperature region we measured.

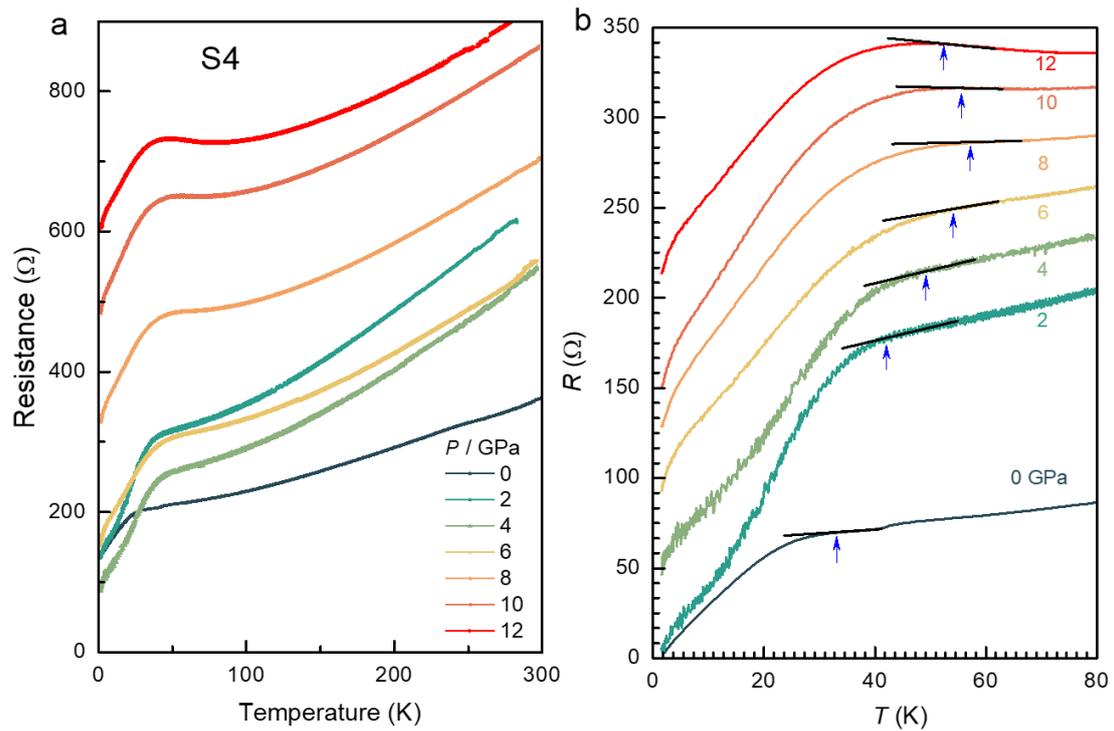

**Extended Data Fig. 7 | Temperature dependence of resistance of S4 under different pressures. a** Temperature dependence of resistance $R(T)$ of a LSP$_{327}$ thin film (S4) under various pressures up to 12 GPa with Daphne 7373 as PTM. **b**, The enlarged view of $R(T)$ curves below 80 K for S4, respectively. To illustrate the variation of the superconducting transition temperatures with pressure more clearly, the $R(T)$ curves in b have been vertically shifted. The $T_c^{onset}$ (up-pointing arrow) was determined as the temperature where resistance starts to deviate from the extrapolated normal-state behavior.

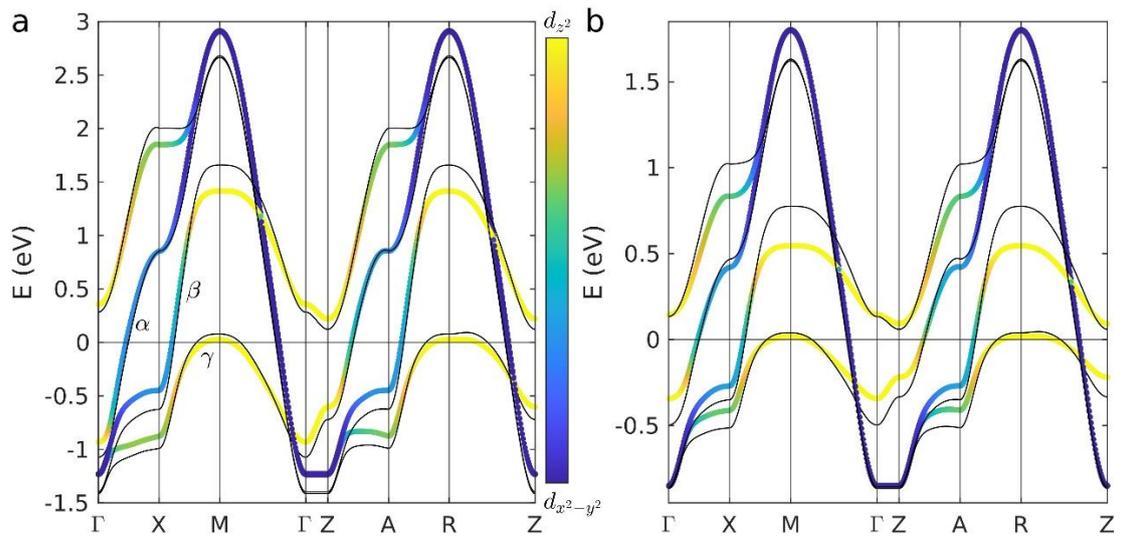

**Extended Data Fig. 8 | DFT results by using different $c$ lattice constants.** Calculated DFT **a** and slave-boson renormalized **b** band structures for the two-orbital bilayer model for ambient pressure shown with color scheme visualizing the orbital content and for $\Delta c/c_0$ = 5% shown as black line.

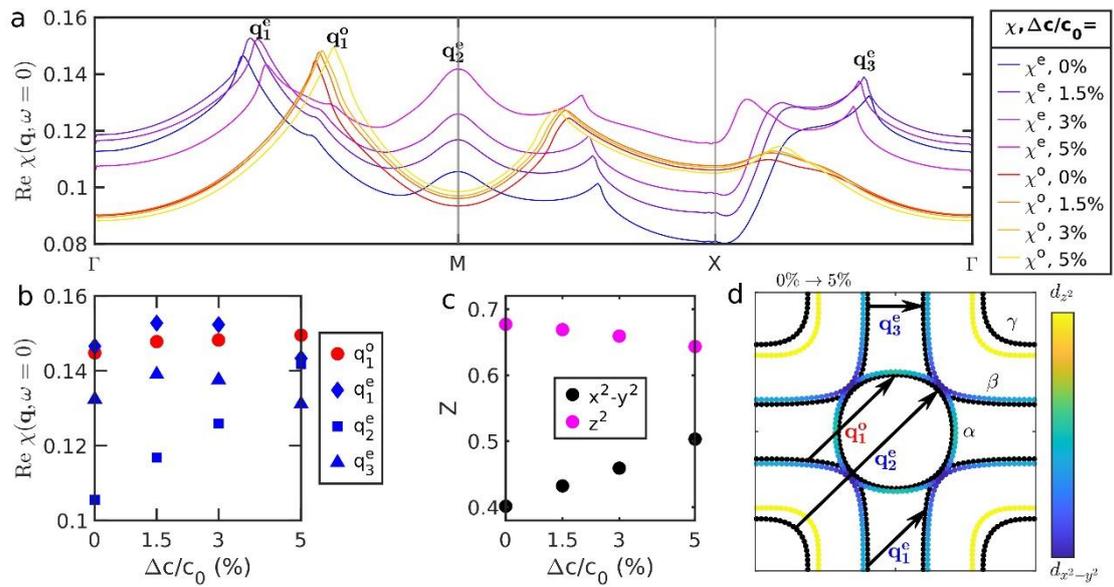

**Extended Data Fig. 9 | Theoretical evolution of the spin fluctuations upon pressure.** Same as Figure 5 but for a doping of 0.1 holes per Ni atom.

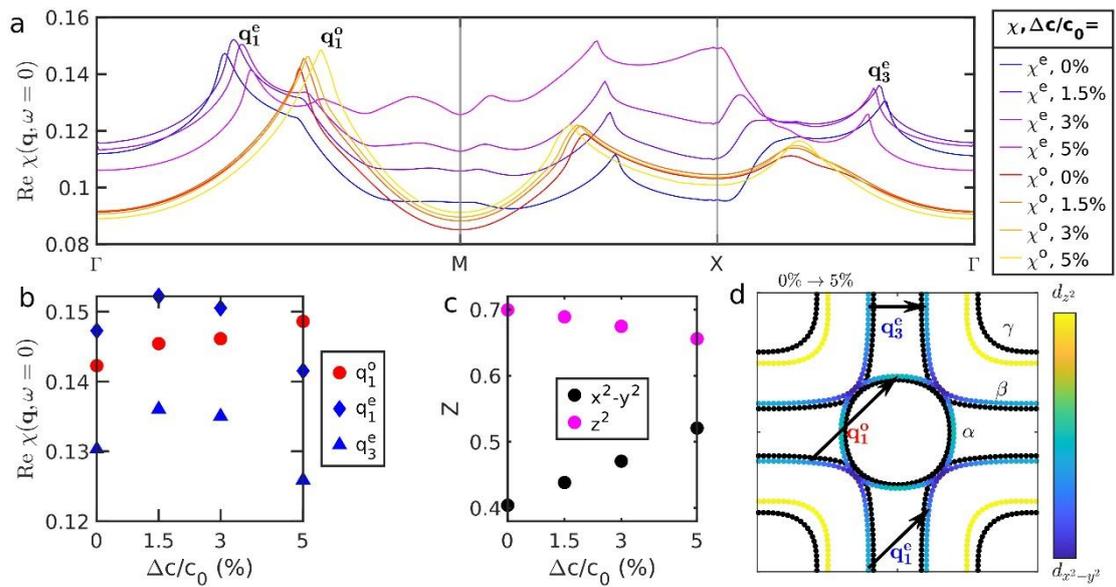

**Supplementary Fig. 10 | Theoretical evolution of the spin fluctuations upon pressure.** Same as Figure 5 but for a doping of 0.2 holes per Ni atom.